\documentclass[%
 reprint,
 showpacs,
 prl,
 amsmath,amssymb,
 aps,
]{revtex4-1}

\usepackage{graphicx}
\usepackage{dcolumn}
\usepackage{bm}

\begin{document}

\preprint{APS/123-QED}

\title{Chaotic properties of a turbulent isotropic fluid}

\author{Arjun Berera}
 \email{ab@ph.ed.ac.uk}
\author{Richard D. J. G. Ho}%
 \email{richard.ho@ed.ac.uk}
\affiliation{%
SUPA, School of Physics and Astronomy, University of Edinburgh,
JCMB, King's Buildings, 
Peter Guthrie Tait Road EH9 3FD, Edinburgh, United Kingdom.
}%

\date{\today}

\begin{abstract}
By tracking the divergence of two initially close trajectories in phase space
in an Eulerian approach to forced turbulence, the relation between the 
maximal Lyapunov exponent $\lambda$, and the Reynolds number $Re$
is measured using direct numerical simulations,
performed on up to $2048^3$ collocation points.
The 
Lyapunov exponent is found to solely depend on the Reynolds number with
$\lambda \propto Re^{0.53}$ and that after a
transient period
the divergence of trajectories grows at the same rate 
at all scales. Finally a linear divergence is seen that is dependent
on the energy forcing rate.
Links are made with other chaotic systems.

\textit{In Press Physical Review Letters 2018}
 
\end{abstract}

\pacs{47.27.Gs, 05.45.-a, 47.27.ek}

\maketitle



Turbulence displays chaotic dynamics
\cite{BohrBook} and 
ideas from chaos theory find many different applications in
turbulence including the dispersion of pairs of particles 
\cite{Taylor1921,Richardson1926,Salazar2009,Biferale2005}, the presence of 
Lagrangian coherent structures 
\cite{Haller2015}, turbulent mixing \cite{Ottino1990}, turbulent transitions 
\cite{Eckhardt2007} and predictability 
\cite{Lorenz1963,*Lorenz1969,Aurell1997,Leith1971,Leith1972,Boffetta2002}. 
Chaos has been seen and applied in systems as diverse as 
quantum entanglement, where the classical dynamical properties are linked
to the quantum counterparts \cite{Jalabert2001,Jacquod2009},
planetary dynamics \cite{Laskar1990},
and biological systems \cite{May1974}.

Using the Eulerian approach,
we track the divergence of fluid field trajectories, which
initially differ by a small perturbation.
We do a model independent analysis, evolving the Navier-Stokes
equations for 
three dimensional homogeneous isotropic turbulence (HIT)
using direct numerical simulation (DNS).
The Eulerian approach to the study of the chaotic properties of turbulence
has received only limited numerical tests prior to this Letter.
Amongst approximate models, there have been
EDQNM closure approximations \cite{Metais1986} 
and shell model studies
\cite{Crisanti1993,Aurell1996,Yamada2007}.
Amongst exact DNS studies, there have been some in two
dimensions \cite{Kida1990,Boffetta1997,Boffetta2001} and
single runs in three dimensions at comparatively small box
sizes \cite{Deissler1986,Kida1992}, all more than a decade
and a half ago.
This Letter tests the 
theory of Ruelle \cite{Ruelle1979}
relating the maximal Lyapunov exponent $\lambda$ and $Re$
in DNS of HIT in a Eulerian sense.
The paper
also examines the time history of the divergence and
finds a uniform exponential growth rate across all
scales at an intermediate time and to show a linear growth for late
time in three dimensional HIT.
The simulations are also the largest yet for measuring the Eulerian
aspects of chaos in HIT for DNS,
performed on up to $2048^3$ collocation
points and reach an integral scale Reynolds number of $6200$.
This allows a more accurate measurement of the $Re$
dependence of $\lambda$.

For a chaotic system, an initially small perturbation  $|\delta \bm{u}_0|$
should grow according to $|\delta \bm{u} (t)| \simeq 
|\delta \bm{u}_0| e^{\lambda t}$
where $t$ is time.
It is theoretically predicted that the Lyapunov exponent should
depend on the Reynolds number according to the
rule \cite{Ruelle1979,Crisanti1993b}
\begin{equation}
 \lambda \sim \frac{1}{\tau} \sim \frac{1}{T_0} Re^{\alpha} \ , 
\ \alpha = \frac{1-h}{1+h} \ .
\label{eq:two}
\end{equation}
The Holder exponent, $h$, is
given by $|\bm{u}(\bm{x} + \bm{r}) - \bm{u}(\bm{x})| 
\sim Vl^h$, where $V$ is the rms velocity, 
$l$ the size of the eddy,
$Re = VL/\nu$
the integral scale Reynolds number, 
$L = (3 \pi / 4 E) \int (E(k)/k)dk $ the integral length scale,
$E$ the energy,
$\nu$ the viscosity,
$T_0 = L/V$ the large eddy turnover time,
$\tau = (\nu/\epsilon)^{1/2}$ the Kolmogorov time scale,
and $\epsilon$ the dissipation rate.
In the Kolmogorov theory, $h$ is predicted to be $1/3$ and so
$\alpha$ is predicted to be 
$1/2$ \cite{Ruelle1979,Crisanti1993b,Kolmogorov1941}.

Some of the new results found in this Letter from the Eulerian approach
are inaccessible to the Lagrangian approach, such as the linear
growth rate of the divergence at late times which has no direct Lagrangian
counterpart.
The paper also
highlights different results from the two approaches.
For instance,
within the Lagrangian approach,
the relation $\lambda \tau \approx const$ has been found
before in tracer particles \cite{Biferale2005,Bec2006} and
for infinitesimal volume deformation \cite{Girimaji1990}.
Furthermore, these results suggest that $\lambda \tau$
decreases slightly with Reynolds number, and that due to intermittency
corrections this implies $\alpha < 0.5$ \cite{Crisanti1993,Bec2006}. 
As will be shown, we find that in the Eulerian approach
$\lambda \tau$ increases slightly with Reynolds
number, which is consistent with our result that $\alpha > 0.5$.
There is nothing which says the Lyapunov exponent in the
Eulerian and Lagrangian frames should be the same. 
An example is ABC flow in which the Lyapunov exponent in the
Lagrangian frame is positive but in the Eulerian 
frame is non-positive \cite{Dombre1986}.
The prediction of Ruelle for turbulence does not distinguish between
Eulerian and Lagrangian frames \cite{Ruelle1979}.


We perform DNS of forced HIT on the 
incompressible Navier-Stokes equations using a fully de-aliased
pseudo-spectral code in a periodic cube of length $2\pi$
\begin{equation}
 \partial_t \bm{u} = - \nabla P - \bm{u} \cdot \nabla \bm{u} +
 \nu \Delta \bm{u} + \bm{f} \ , \ \nabla \cdot \bm{u} = 0 \ , 
\end{equation}
where $\bm{u}$ is the velocity field, $P$ the pressure, $\nu$ 
the viscosity and $\bm{f}$ the external forcing. The density was 
set to unity \cite{TheData2017}. 
The primary forcing used was a negative damping scheme which 
only forced the low wavenumbers (large scales), 
$k_f = 2.5$, according to the rule
\begin{equation}
\bm{\hat{f}}(\bm{k},t) = 
\begin{array}{c}
(\epsilon / 2E_f) \bm{u}(\bm{k},t) \ \text{if} \ 0 < |\bm{k}| < k_f ; \\
0 \ \text{otherwise}
\end{array}  \ ,
\label{eq:forcing}
\end{equation}
where $E_f$ is the energy in the forcing band
and $\bm{u}(\bm{k},t)$ is the Fourier coefficient of field $\bm{u}$.
This well tested forcing function \cite{Linkmann2015,Kaneda2006}
allows the dissipation rate, $\epsilon$, to be known a priori.
We set $\epsilon$ to 0.1 for all runs unless otherwise stated. 
A full description of the code, including the forcing, 
can be found in \cite{YoffeThesis}. The Reynolds number quoted 
throughout this Letter is the integral 
scale Reynolds number, $Re$, which
was changed by varying $\nu$. 
The simulations were well resolved, with $k_{max} \eta > 1$ for
all simulations, where $k_{max}$ is the largest wavenumber in the 
simulation and $\eta$ the Kolmogorov length. 
$T_0$ and $L$ vary between simulations.
Over resolved simulations, with $k_{max} \eta \gg 1$, were performed 
to test if the box size
had a statistically significant effect on the results and this was not
the case.
All simulations parameters are given in the Supplementary Material.

To implement the perturbation, a copy of the evolved field $\bm{u}_1$ 
was made and perturbed slightly to create field $\bm{u}_2$. 
This perturbation was achieved by not calling the forcing function 
at one particular timestep.
This meant that the perturbation would be in the band of wavenumbers
$0 < |\bm{k}| < k_f$ and would depend non-trivially on the field itself
by Eq.~(\ref{eq:forcing}).
The difference field $\delta \bm{u} = 
\bm{u}_1 - \bm{u}_2$ was then calculated. Fields $\bm{u}_1$ and 
$\bm{u}_2$ were then evolved independently and the statistics of 
$\delta \bm{u}$ were tracked. 
The same realisation of the external forcing is used on
both fields.
The key statistic measured was the energy 
spectrum of the field, 
$E(k,t)$, which in Fourier space is defined by
\begin{equation}
E(k,t) = \frac{1}{2} \int_{|\bm{k}|=k} d 
\bm{k} |\bm{\hat{u}}( \bm{k},t) |^2 \ ,
\end{equation}
with total energy, $E(t) = \int_0^\infty dk E(k,t)$. Analogously, 
we define the energy of the difference spectrum, $E_d(k,t)$ as
\begin{equation}
E_d(k,t) = \frac{1}{2} \int_{|\bm{k}|=k} d \bm{k} 
|\bm{\hat{u}}_1 ( \bm{k},t) -  \bm{\hat{u}}_2 ( \bm{k},t)|^2 \ ,
\label{eq:one}
\end{equation}
which is useful in assessing the degree of divergence of two fields at a 
particular scale. We then similarly define
 $E_d(t) = \int_0^\infty dk E_d(k,t)$ as the total 
energy in the difference spectrum. By inspection we can see 
that $|\delta \bm{u} (t)| = (2E_d(t))^{1/2}$.

\begin{figure}
\includegraphics[width=0.5\textwidth]{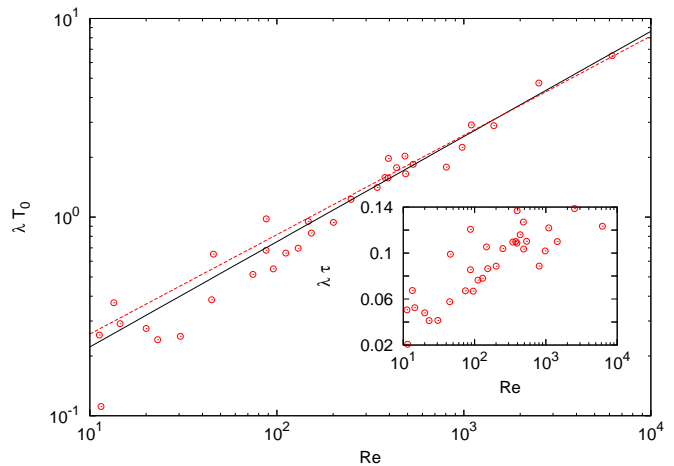}
\caption{\label{fig:plot2} The main plot shows $Re$ against $\lambda T_0$ 
and the fit 0.066$Re^{0.53}$ as a solid black line.
Errors for the higher wavenumbers are
comparable to the size of the points and are not included for clarity.
The lower wavenumbers have larger error.
A line of $Re^{0.5}$ fit to the data is shown in dashed red (gray).
The inset shows $\lambda \tau$ against $Re$ for the same data.}
\end{figure}

After a statistically steady state of 
turbulence was reached, perturbations were made for a 
range of Reynolds numbers from $Re \approx 10$ 
to $Re \approx 6200$ at box sizes
from $64^3$ to $2048^3$. We found that the
growth of $|\delta \bm{u}|$ best fit an exponential $\exp(\lambda t)$. 
We multiply $\lambda$ by $T_0$ to non-dimensionalize
the simulation time. 
A plot of $Re$ vs. $\lambda T_0$ is shown in 
Fig.~\ref{fig:plot2}. From the data we find a good fit to the functional
form $\lambda T_0 = C Re^\alpha$ 
with $\alpha = 0.53 \pm 0.03$ and constant $C = 0.066 \pm 0.008$,
in reasonable agreement with the theory value prediction
\cite{Ruelle1979}. 
Previous results from a shell model analysis relying on a
phenomenological multifractal model to extract a fit gave
$\alpha = 0.459$ \cite{Crisanti1993},
whilst other Lagrangian results have suggested
$\alpha < 0.5$ \cite{Bec2006}.
We cross-checked the $Re$ dependence using an 
alternative DNS implementation of HIT described in \cite{Chumakov2008},
which gave a result within one standard error of ours (see Supplementary
Material).
In a Lagrangian study \cite{Bec2006} 
a decrease in $\lambda \tau$ was associated
with $\alpha < 0.5$. As is shown in the inset in Fig.~\ref{fig:plot2}
our data shows an increase in $\lambda \tau$ with $Re$, which
agrees with $\alpha > 0.5$ found here. 
This shows at least one difference between the Eulerian and Lagrangian
approaches, which may have some significant underlying reason
worth exploring in future work.

\begin{figure}
\includegraphics[width=0.5\textwidth]{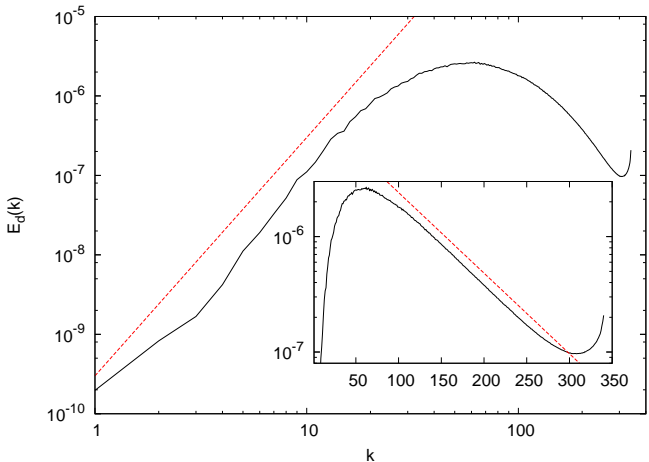}
\caption{\label{fig:plotEdk} $E_d(k)$ in black at an intermediate
time for a simulation with  $Re \approx 2500$ on box size $1024^3$,
the main plot is logarithmic and has dashed red (gray)
line showing $k^3$ whilst the inset is semi-logarithmic with dashed red (gray)
line showing an exponential slope.}
\end{figure}

\begin{figure}
\includegraphics[width=0.5\textwidth]{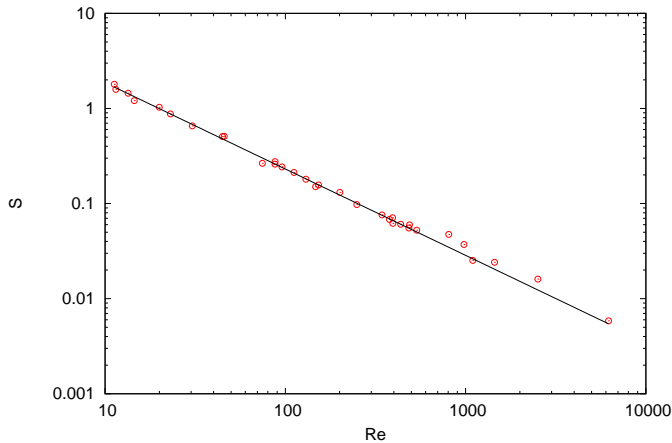}
\caption{\label{fig:plotslope} $|S|$ vs. $Re$ with fit
$15.1*Re^{-0.91}$, where high-k behavior of $E_d(k)$ is
approximated as $\exp(-Sk)$.}
\end{figure}

We find that an initial perturbation must adopt a particular spectrum,
described below for $E_d(k)$, before $E_d(k)$ grows uniformly at all scales
and maintains this profile during exponential growth.
This particular spectrum is shown in Fig.~\ref{fig:plotEdk} for a run with
$Re \approx 2500$.
The spectrum of $E_d(k)$ has three main 
characteristics; at low $k$ there is an approximately $k^3$ power
law dependence, at intermediate $k$ $E_d(k)$ has a peak between 
the peaks of $E(k)k^2$ and
$E(k)k^3$, and for high $k$ there is an exponential dependence on wavenumber,
which we approximate as $E_d(k) \sim \exp(-Sk)$.
Our DNS show that this exponential slope becomes flatter with increasing
$Re$ according to a power law, this dependence is very strong and is shown
in Fig.~\ref{fig:plotslope} which plots the relationship between $Re$ and
the magnitude of the exponential slope, $S$.
Thus, as $Re$ becomes large, 
$E_d(k)$ becomes flat for wavenumbers higher than the peak.
The difference spectrum at low $k$ for an EDQNM approximation was found to
be $k^4$ \cite{Metais1986}, whilst in a single run of DNS it was $k^2$ with
large error \cite{Kida1992}.
Similar difference spectra as ours at all scales
have been seen in atmospheric models \cite{Vannitsem2017}.

To understand the origin of the peak in $E_d(k)$, it is
useful to look at the theory of \cite{Ruelle1979}, where
it is assumed that the maximal Lyapunov
exponent is inversely proportional to the smallest characteristic eddy time,
which is the Kolmogorov time $\tau$. 
Naively we might expect that the peak of $E_d(k)$ to be $k_\eta$, the
wavenumber corresponding to $\eta$, which is the Kolmogorov length scale with
$\eta = (\nu^3/\epsilon)^{0.25}$. This is not observed.
Instead, we can define
a frequency for eddies at wavenumber $k$ of $f(k)k$ where 
$f(k) = \sqrt{E(k)k}$ \cite{HinzeBook}.
This would make the divergence dominated by the
eddies of the size of the peak of $E(k)k^3$, which is close to the
observed peak of $E_d(k)$.

\begin{figure}
\includegraphics[width=0.5\textwidth]{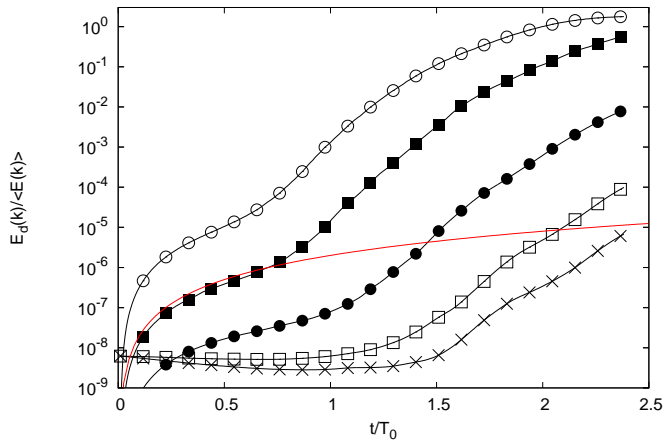}
\caption{\label{fig:plot4} (color online)
The growth of $E_d(k)/\langle E(k)\rangle$
for selected wavenumbers with time for a simulation with $Re \approx 2500$
and box size $1024^3$. The wavenumber increases upwards, the plotted
wavenumbers are $k = 1,2,5,20,100$,
in turn these are represented by crosses, empty squares, solid circles,
solid squares, and empty circles.
The red (gray) line follows wavenumber $k=20$ and shows the $t^2$ 
dependence for early times.
The perturbation was performed at low-k, at all wavenumbers between
0 and 2.5.}
\end{figure}

It is also interesting to plot the growth of 
$E_d(k)/\langle E(k)\rangle$ for selected wavenumbers 
as is done in Fig.~\ref{fig:plot4}, for the run with $Re \approx 2500$
on box size $1024^3$, with angled brackets representing a steady state
average.
The perturbation was performed at the forcing wavenumbers, $k < k_f$.
There are three stages of growth.
The first stage is a transient stage during which the characteristic
$E_d(k)$ spectra is adopted. For the low wavenumber perturbation,
the large scales remain close for at least one $T_0$, waiting until
the small scale divergence has reached a certain size, as
seen before in one dimensional atmospheric models \cite{Lorenz1996}.
This is the cause for the different behaviour of $k = 1,2$ in
Fig.~\ref{fig:plot4} compared to the other wavenumbers.
In our simulations $E_d(k) \sim t^2$ for the small scales when
the perturbation was made at low wavenumber.
If the perturbation is made at high wavenumber, the large scales do not
remain close and there is an initial convergence of the fields,
as seen in 2D turbulence, suggesting a common behaviour \cite{Boffetta1997}.
If the perturbation is made at low wavenumber then there is no initial
convergence.

Note that, although the plot in Fig.~\ref{fig:plot4} is of
one particular initial state and initial perturbation vector, 
we find that
the presence of these three stages appears to be independent of the form
of the perturbation made and initial state. 
Only the initial transient stage depends on
the form of the perturbation. Perturbations made at high wavenumber
exhibited the same form in the latter two stages as those made at low
wavenumber. This suggests it is a characteristic feature of the difference
field evolution.

The second stage is the exponential growth stage, where it is notable that all
scales grow at the same exponential rate and this exponent is the
same as the maximal Lyapunov exponent.
In test simulations, 
forcing was performed at intermediate wavenumbers so that wavenumbers lower
than the inertial range could be simulated. These simulations also showed
the same exponential growth rate at every scale, including those
larger than the forcing scale.
This suggests it is not a feature
of the well known forward cascade of energy in turbulence.
This scale independent growth has also been seen in 
quasi-geostrophic turbulence in a channel \cite{McWilliams1981},
atmospheric models \cite{Vannitsem1997,Vannitsem2017}, and other
systems of non-linear equations \cite{Gao2006,Bohr1989}.
We now also measure it in a large turbulent simulation.
In Fig.~\ref{fig:plot4} this stage is relatively short but can
be extended arbitrarily by having a smaller perturbation,
these checks also showed our perturbation could be considered infinitesimal.

\begin{figure}
\includegraphics[width=0.5\textwidth]{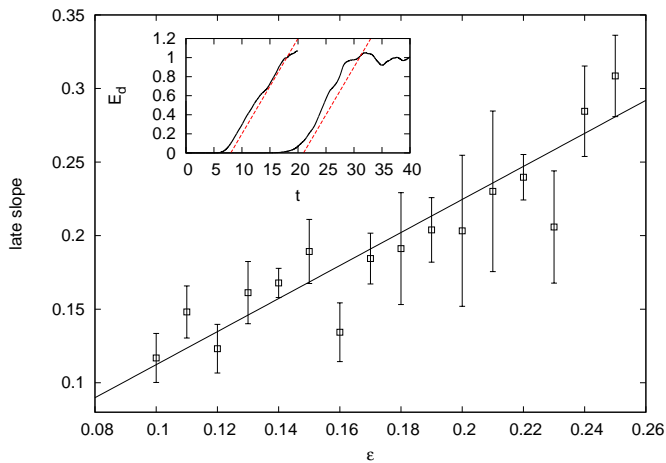}
\caption{\label{fig:plot5} (color online)
The main plot shows $\epsilon$ against $dE_d(t)/dt$ at late times
before saturation, with error shown on measured slope and fit $1.12\epsilon$.
The inset shows $E_d$ for a run with $Re \approx 800$ on the left
and $Re \approx 130$ on the right in black with dashed red (gray)
line with slope $\epsilon$.
}
\end{figure}

The third stage is the late time saturation stage, the details 
of which depend on the size of the inertial range.
At late times, the growth of $E_d(t)$ enters 
a linear stage before saturation, which is
entered as soon as $dE_d(t)/dt \approx \epsilon$.
This implies that the
threshold energy is $E_d \approx \epsilon/
2 \lambda$. If this energy
is greater than the saturation of the difference,
then the growth of the difference is exponential until it saturates.
$E_d$ for runs at $Re \approx 130$ and
$Re \approx 800$ are shown in the inset
of Fig.~\ref{fig:plot5}, where late time starts at $t \approx 20$ for
$Re \approx 130$ and $t \approx 7$ for $Re \approx 800$.

By varying the rate of dissipation
we can see the dependence of this
linear growth rate on $\epsilon$, which is the energy input rate
for a statistically steady state system. A plot of $\epsilon$ against
$dE_d(t)/dt$ for late times is shown in Fig.~\ref{fig:plot5}.
The values here are not normalized and we find
$dE_d(t)/dt = 1.12 \epsilon$.
$dE_d(t)/dt$ is really a quantification of the 
rate of separation of trajectories
in phase space, which is related to information creation, i.e.
Kolmogorov-Sinai (KS) entropy.
If it is possible to interpret $dE_d(t)/dt$ as the KS
entropy, we can relate our results with 
corollary (2.2) of \cite{Ruelle1982} 
which shows that the upper bound of the KS entropy in an isothermal fluid
in equation (2.9) of \cite{Ruelle1982} is related to the dissipation.

The findings of linear growth in $E_d$ at late time in a 2D DNS of turbulence
were justified on the basis that there is a characteristic timescale
for the eddies $\tau(k) \sim k^{-2/3}$ \cite{Boffetta2001}, 
which is in agreement with the definition of our frequency $f(k)k$.
However, in our data we find
instead that $\tau(k) \sim k^{-1/3}$.
This linear growth at late times does not have a clear Lagrangian
counterpart.
For high $Re$ the exponential growth phase may be very brief and
so the majority of the divergence will be dominated by the linear growth,
which only depends on the dissipation. In this way the divergence of two
velocity field trajectories
may be universal in the Kolmogorov sense at high $Re$.


We have found that, if one scale $E_d(k)$ diverges exponentially, then all
scales do so. This could indicate the presence of a turbulent regime.
If there is no turbulent regime, then there are no scales which diverge
exponentially in the Eulerian framework.
This is different to the Lagrangian case.
Instead of associating the inverse Lyapunov exponent with
Kolmogorov time $\tau$, a slight reinterpretation
of Ruelle's theory is to associate the characteristic time with
$l_T/V$ where $l_T$ is the Taylor microscale,
which only exists
if an inertial range exists (see Supplementary Material for data). 
This would also give $\alpha$ close to 0.5.
This quantity uses the largest velocity and smallest length scale
exclusive to turbulence 
to achieve the smallest time scale.

In summary, we have shown that the degree of chaos for forced 
HIT appears to be uniquely 
dependent on the large scale Reynolds number according to the law 
$\lambda T_0 \sim Re^{0.53}$. Divergence does not occur
at all scales until the velocity field difference spectrum
adopts a characteristic form.
After this spectrum is adopted, the normalized energy difference spectrum 
$E_d(k)/\langle E(k)\rangle$ grows similarly for all wavenumbers at
intermediate times.
Due to the shape of the spectrum, the smallest length scales will become 
decorrelated long before the largest length scales, as 
has been predicted before \cite{Lorenz1969}. 
At the large scales, predictability for a fixed tolerance 
should be possible for much longer than at the smallest scales.
The late time growth of $E_d(t)$ was found to be linear and approximately
equal to the energy input rate.

This Letter has made thorough 
numerical demonstrations of the links between
chaos and turbulence in a Eulerian context, 
and so by extension relates turbulence
to other chaotic processes and might provide a different perspective for
their study.
In chaos containing multiple length and time scales, 
applying ideas from turbulence may be especially fruitful because
we have seen similar features here in turbulence to those
found in chaotic systems which are not considered turbulent
\cite{Kandrup2003,Vannitsem2017,Bohr1989,Gao2006}.
There are interesting similarities between the linear growth behavior found
in this paper and others \cite{Ruelle1982,Bianchi2017}, which we will
examine in the future.

\begin{acknowledgments}
We would like to thank Moritz Linkmann for initial help with the project
and further useful input and discussion.
We would also like to thank Sergei Chumakov for help with, and provision
of, the alternative
DNS code (\url{https://code.google.com/archive/p/hit3d}).
This work has used resources from the Edinburgh Compute and Data Facility
(\url{http://www.ecdf.ed.ac.uk})
and ARCHER (\url{http://www.archer.ac.uk}). A.B acknowledges support from 
the UK Science and Technology Facilities Council
whilst R.D.J.G.H is supported by the UK Engineering and Physical Sciences
Research Council (EP/M506515/1).
\end{acknowledgments}


\bibliographystyle{apsrev4-1}
\providecommand{\noopsort}[1]{}\providecommand{\singleletter}[1]{#1}%

\end{document}


\preprint{APS/123-QED}

\title{Chaotic properties of a turbulent isotropic fluid: supplementary material}

\author{Arjun Berera}
\author{Richard D. J. G. Ho}%
\affiliation{%
SUPA, School of Physics and Astronomy, University of Edinburgh,
JCMB, King's Buildings,
Peter Guthrie Tait Road EH9 3FD, Edinburgh, United Kingdom.
}%

\date{\today}

\begin{abstract}

\end{abstract}

\maketitle

\section{The response to perturbation}

\begin{figure}
\includegraphics[width=0.5\textwidth]{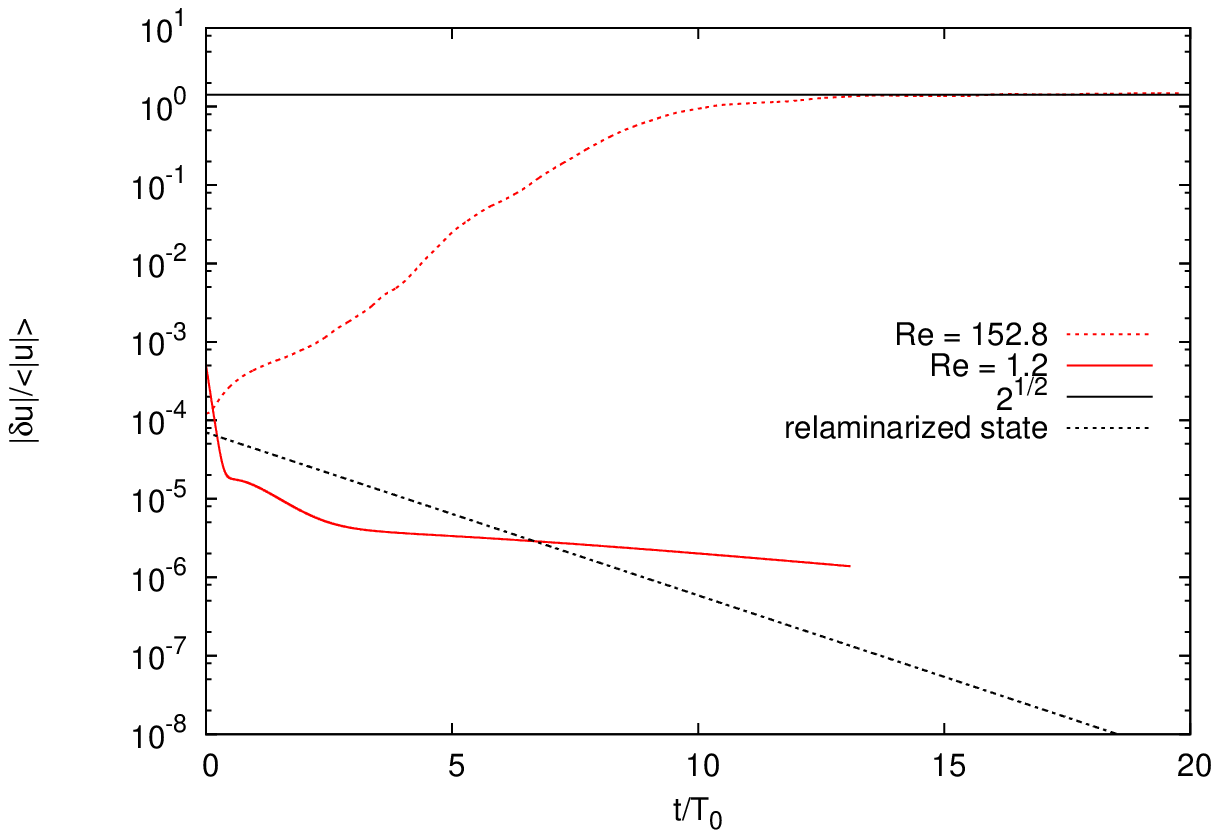}
\caption{\label{fig:plot1} The evolution of
$| \delta \bm{u}| / \langle |\bm{u}| \rangle$ as a function
of dimensionless time $t/T_0$ for $Re \approx 150$ and in dashed red,
$Re \approx 1.2$ in solid red and a relaminarized state in dashed black
as well as the saturation prediction in solid black.}
\end{figure}

Fig.~\ref{fig:plot1} shows the evolution of
$| \delta \bm{u}| / \langle |\bm{u}| \rangle$, which demonstrates
the behaviour of the divergence of the two fields.
We normalize $| \delta \bm{u} | / \langle |\bm{u}| \rangle$
by the steady state value $\langle |\bm{u}| \rangle$ for convenient comparison.
The response of a higher $Re$ run ($Re \approx 150$ and $N=128$),
a low $Re$ run ($Re \approx 1.2$ and $N=64$),
and a relaminarized state ($Re \approx 23$ and $N=64$)
\cite{Linkmann2015} are compared.
These relaminarized states are akin to those found in
parallel wall-bounded shear flows \cite{Linkmann2015}.
For the higher $Re$ run there is a clear exponential growth of
$| \delta \bm{u} | / \langle |\bm{u}| \rangle$ with a
slight levelling off before saturation at $\sqrt{2}$,
which is predicted by assuming
that $\int \bm{u}_1 \cdot \bm{u}_2 d \bm{x} = 0$ over all space.
The low $Re$ run, perturbed before any relaminarization occurred,
shows a convergence. This is qualitatively
different from the relaminarized state, which
shows an exponential convergence of the
perturbed field, thus confirming the linear stability
of the relaminarized state as suggested in \cite{Linkmann2015}.
As such, the behavior in the three separate regimes is consistent
with expectations.

\section{hit3d code results}

\begin{figure}[h!]
\centering
\includegraphics[width=0.5\textwidth]{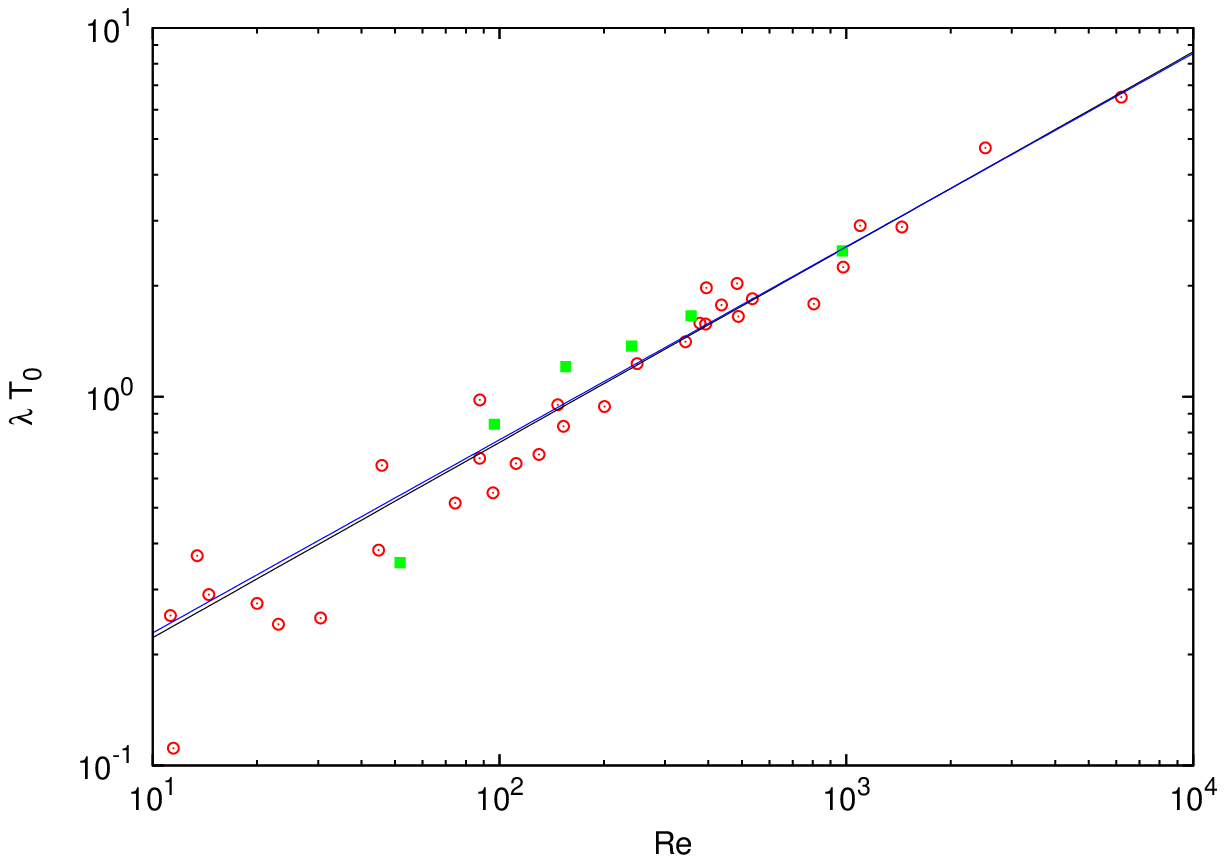}
\caption{\label{fig:plothit3d} $Re$ against $\lambda T_0$, results from our own code shown as red circles, results from hit3d code as green squares. The black line shows $0.066*Re^{0.53}$, the dark blue line shows $0.068*Re^{0.53}$.}
\end{figure}

An alternative implementation of homogeneous isotropic turbulence was used to test the results, this was the hit3d code developed by Sergei Chumakov (\url{https://code.google.com/archive/p/hit3d}). The results from this code for the relationship between $Re$ and $\lambda T_0$ is shown in Fig.~\ref{fig:plothit3d}

In Fig.~\ref{fig:plothit3d} we show both the fit to all the data and the fit to only the runs with $Re > 400$, this is fitting the function $c*Re^{\alpha}$. The fit with all the data gives $c = 0.066 \pm 0.008$ and $\alpha = 0.53 \pm 0.03$ whilst the fit with only $Re > 400$ gives $c = 0.068 \pm 0.019$ and $\alpha = 0.53 \pm 0.06$. In the figure, these two fits, though both plotted, are nearly indistinguishable.

\clearpage

\section{Alternative characteristic timescale and Lyapunov exponent}

\begin{figure}[h!]
\centering
\includegraphics[width=0.5\textwidth]{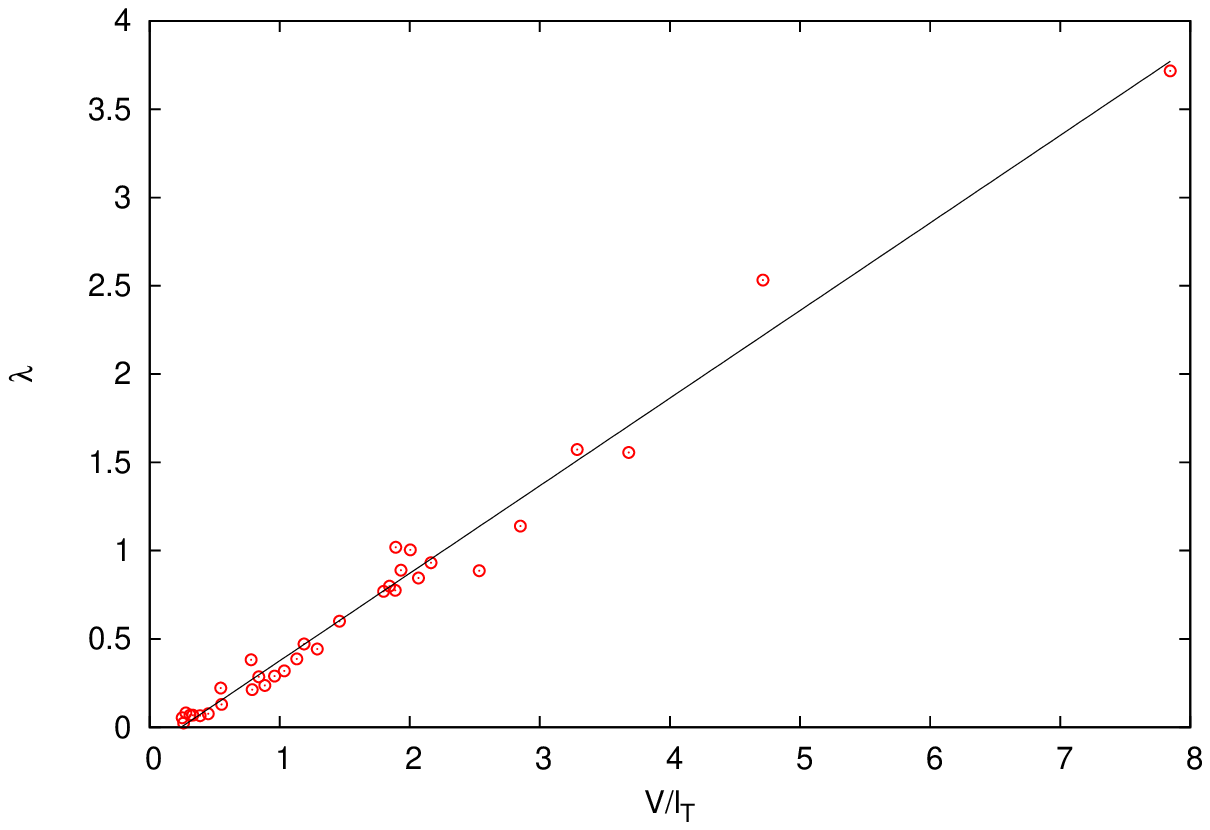}
\caption{\label{fig:plotscales} $V/l_T$ against $\lambda$, the solid line is given by $\lambda = 0.496 (V/l_T) - 0.12$.}
\end{figure}

The reinterpretation of the characteristic timescale for chaos as $l_T/V$ where $l_T$ is the Taylor microscale and $V$ would mean that $\lambda \propto V/l_T$. A graph of these two quantities is presented in Fig.~\ref{fig:plotscales}. The linear relationship is given by $\lambda = m*(V/l_T) + c$ where $m = 0.496 \pm 0.012$ and $c = -0.12 \pm 0.03$.

\section{Saturation of $E_d$ and when $\partial_t E_d$ is maximised}

The difference between the two fields will eventually saturate because the fields do not have infinite energy. We may predict a saturation point for this measure of the difference field as $\sqrt{2}$. If we assume $|\bm{u}_1| = |\bm{u}_2| = |\bm{u}_2 - \delta \bm{u}|$
\begin{align}
\sum_i u_{1i}^2 &= \sum_i (u_{2i} + \delta u_i)^2 = \sum_i (u_{2i}^2 + 2 u_{2i} \delta u_i + \delta u_i^2) \\
\sum_i \delta u_i^2 &= -\sum_i 2 u_{2i} \delta u_i 
\end{align}

where $i$ is the index in phase space, and also assume that the final fields are completely uncorrelated
\begin{align}
\sum_i u_{1i} u_{2i} &= 0 = \sum_i u_{2i}( u_{2i} + \delta u_i) \\
\sum_i u_{2i} \delta u_i &= - \sum_i u_{2i}^2 \\
\sum_i \delta u_i^2 &= 2 \sum_i u_{2i}^2
\end{align}

Thus $|\delta \bm{u}| = \sqrt{2}|\bm{u}_1|$ after sufficient time.

If the growth of $E_d$ is limited by the dissipation, we can predict the value of $E_d$ at which the time it is limited by dissipation occurs
\begin{align}
E_d(t) &= d_0 e^{2 \lambda t} \\
\frac{d E_d(t)}{dt} &= 2 \lambda d_0 e^{2 \lambda t}
\end{align}

where $d_0$ is the initial separation and $\lambda$ is the separation rate. The factor of 2 comes from the normalisation. When we are limited by dissipation we have
\begin{align}
\frac{d E_d(t)}{dt} = \epsilon = 2 \lambda d_0 e^{2 \lambda t} = 2 \lambda E_d
\end{align}

a simple rearrangement gives $E_d = \epsilon / 2 \lambda$.

\section{Description of alternative forcings}

In order to test the robustness of the results, we tried alternative
forcing strategies to the negative damping described in the main text.
These were a static sine forcing and an adjustable helicity forcing, which
had a stochastic element, both described in \cite{Linkmann2017}.

We can define the external forcing of field $\bm{u}_1$ to be $\bm{f}_1$ and
of field $\bm{u}_2$ to be $\bm{f}_2$.
We find that if $\bm{f}_1 \approx \bm{f}_2$, (such as initially for the
negative damping, the sine forcing, or as possible with the adjustable
helicity forcing) then we find that the three phases of growth (initial
transient, exponential growth, and linear growth) proceed as described
before.

However, if $\int \bm{f}_1 \cdot \bm{f}_2 d \bm{x} \approx 0$, (as is possible
in the stochastic forcing) then we are stuck in the initial transient stage.
The perturbation grows as a power law with tie. The reason for this is that
effectively
a perturbation is added at every timestep between the two fields $\bm{u}_1$ 
and $\bm{u}_2$ which swamps the exponential growth.

\bibliographystyle{apsrev4-1}
\providecommand{\noopsort}[1]{}\providecommand{\singleletter}[1]{#1}%

\clearpage

\section{Simulation parameters}

\begin{table}[h!]
\centering
\begin{tabular}{ccccccccc}
$N^3$ & $\nu$ & $\tau$ & $Re$ & $Re_l$ & $\lambda$ & $j$ & $T_0$ & $L$ \\
\hline
$2048^3$ & 0.00011 & 0.0332 & 6210 & 453 & 3.718 & -0.00585 & 1.75 & 1.09 \\ 
$1024^3$ & 0.0003 & 0.0548 & 2520 & 286 & 2.533 & -0.0161 & 1.87 & 1.19 \\ 
$1024^3$ & 0.0005 & 0.0707 & 1450 & 212 & 1.556 & -0.0241 & 1.86 & 1.16 \\ 
$512^3$ & 0.0006 & 0.0775 & 1100 & 180 & 1.573 & -0.0253 & 1.85 & 1.1 \\ 
$512^3$ & 0.0008 & 0.0894 & 979 & 174 & 1.139 & -0.0371 & 1.97 & 1.24 \\ 
$512^3$ & 0.001 & 0.1 & 806 & 158 & 0.887 & -0.0474 & 2.01 & 1.27 \\ 
$512^3$ & 0.002 & 0.141 & 344 & 100 & 0.775 & -0.076 & 1.82 & 1.12 \\ 
$512^3$ & 0.005 & 0.224 & 147 & 61.5 & 0.471 & -0.151 & 2.02 & 1.22 \\ 
$512^3$ & 0.01 & 0.316 & 87.7 & 43.8 & 0.381 & -0.276 & 2.57 & 1.5 \\ 
$512^3$ & 0.02 & 0.447 & 45.8 & 28.5 & 0.221 & -0.51 & 2.94 & 1.64 \\ 
$256^3$ & 0.0014 & 0.118 & 536 & 125 & 0.931 & -0.0525 & 1.98 & 1.22 \\ 
$256^3$ & 0.0015 & 0.122 & 488 & 121 & 0.845 & -0.0597 & 1.96 & 1.2 \\ 
$256^3$ & 0.0016 & 0.126 & 484 & 120 & 1.004 & -0.0553 & 2.02 & 1.25 \\ 
$256^3$ & 0.0017 & 0.13 & 436 & 113 & 0.889 & -0.0606 & 1.99 & 1.22 \\ 
$256^3$ & 0.0018 & 0.134 & 395 & 108 & 1.019 & -0.062 & 1.94 & 1.17 \\ 
$256^3$ & 0.0019 & 0.138 & 378 & 103 & 0.798 & -0.0681 & 1.98 & 1.19 \\ 
$256^3$ & 0.002 & 0.141 & 393 & 107 & 0.769 & -0.071 & 2.05 & 1.27 \\ 
$256^3$ & 0.003 & 0.173 & 249 & 83.3 & 0.6 & -0.0976 & 2.05 & 1.24 \\ 
$128^3$ & 0.004 & 0.2 & 201 & 73.4 & 0.443 & -0.131 & 2.12 & 1.31 \\ 
$128^3$ & 0.005 & 0.224 & 153 & 62.8 & 0.386 & -0.157 & 2.15 & 1.28 \\ 
$128^3$ & 0.006 & 0.245 & 130 & 57.4 & 0.319 & -0.181 & 2.19 & 1.31 \\ 
$128^3$ & 0.007 & 0.265 & 112 & 51 & 0.289 & -0.212 & 2.28 & 1.34 \\ 
$128^3$ & 0.008 & 0.283 & 95.8 & 46.7 & 0.236 & -0.243 & 2.32 & 1.33 \\ 
$128^3$ & 0.009 & 0.3 & 87.6 & 43.7 & 0.285 & -0.26 & 2.39 & 1.37 \\ 
$64^3$ & 0.01 & 0.316 & 74.5 & 39 & 0.212 & -0.265 & 2.42 & 1.34 \\ 
$64^3$ & 0.02 & 0.447 & 44.8 & 27.3 & 0.129 & -0.507 & 2.97 & 1.63 \\ 
$64^3$ & 0.03 & 0.548 & 30.5 & 20.4 & 0.0755 & -0.656 & 3.32 & 1.74 \\ 
$64^3$ & 0.04 & 0.632 & 23 & 16.1 & 0.0652 & -0.876 & 3.71 & 1.85 \\ 
$64^3$ & 0.05 & 0.707 & 20 & 14.9 & 0.0679 & -1.03 & 4.05 & 2.01 \\ 
$64^3$ & 0.06 & 0.775 & 14.5 & 10.9 & 0.0677 & -1.209 & 4.29 & 1.93 \\ 
$64^3$ & 0.07 & 0.837 & 13.4 & 10.5 & 0.0806 & -1.445 & 4.6 & 2.08 \\ 
$64^3$ & 0.08 & 0.894 & 11.5 & 9.15 & 0.0231 & -1.588 & 4.81 & 2.1 \\ 
$64^3$ & 0.09 & 0.949 & 11.2 & 9.45 & 0.0533 & -1.798 & 4.78 & 2.2 \\ 
\end{tabular}
\caption{\label{tab:table1}
The simulation parameters for runs to determine the Lyapunov exponents.
$N^3$ refers to the number of collocation points simulated, $\nu$ is
the viscosity, $\tau$ the Kolmogorov time, $Re$ the integral scale Reynolds number,
$Re_l$ the Taylor scale Reynolds number, $\lambda$ the non-normalised
Lyapunov exponent, $j$ the exponential slope of $E_d(k) \sim exp(jk)$ for high $k$, $T_0$ the
large eddy turnover time, and $L$ the integral length scale. Perturbations were made at all wavenumbers between 0 and 2.5.}

\end{table}